\documentstyle[11pt,aaspp4]{article}
\begin{document}

\title{Detection of Earth-like Planets Using Apodized Telescopes}

\author{Peter Nisenson}
\affil{Harvard-Smithsonian Center for Astrophysics\\ Cambridge, MA
02138 \\ e-mail: pnisenson@cfa.harvard.edu}
\author{Costas Papaliolios}
\affil{Harvard-Smithsonian Center for Astrophysics and Harvard University 
Dept of Physics \\ Cambridge, MA
02138 \\ e-mail: cpapaliolios@cfa.harvard.edu}

\begin{abstract}
The mission of NASA's Terrestrial Planet Finder (TPF) is to find
Earth-like planets orbiting other stars and characterize the
atmospheres of these planets using spectroscopy. Because of the
enormous brightness ratio between the star and the reflected light
from the planet, techniques must be found to reduce the brightness of
the star. The current favorite approach to doing this is with
interferometry: interfering the light from two or more separated
telescopes with a $\pi$ phase shift, nulling out the starlight.
While this technique can, in principle, achieve the required dynamic
range, building a space interferometer that has the necessary
characteristics poses immense technical difficulties. In this paper,
we suggest a much simpler approach to achieving the required dynamic
range. By simply adjusting the transmissive shape of a telescope
aperture, the intensity in large regions around the stellar image can
be reduced nearly to zero.  This approach could lead to construction
of a TPF using conventional technologies, requiring space optics on a
much smaller scale than the current TPF approach.

\end{abstract}

Subject Headings: Astronomical Instrumentation --- instrumentation:
interferometers --- techniques: interferometric --- Stars --- stars:
imaging --- stars: planetary systems

\section{Introduction}
The mission of the Terrestrial Planet Finder (TPF) is to find
Earth-like planets orbiting other stars by imaging, and to determine
their atmospheric constituents by spectral analysis of their reflected
light (Beichman 1999). Imaging provides separation of the light
reflected from the planet from that of the star. Because the star is
so bright compared to the planet, the dynamic range required is of
order $10^9$ at visible wavelengths. The ``book'' approach proposed
for TPF is to operate in the infrared (7 to 14 microns) where
the dynamic range required is ``only'' $10^6$. However, working in the
IR increases the resolution requirement of the telescope or
interferometer needed to separate the image of the planet from the
image of the star. A 20 meter interferometer baseline or telescope
diameter may be sufficient in the visible band, but in the IR, a 200
meter telescope or 200 meter baseline interferometer is required. Such
a large telescope is out of the question, and a 200 meter
interferometer cannot be built on a single structure, so free flyers
are needed, making for very difficult construction and operation. The
proposed interferometer would be used in a mode where the starlight is
``nulled'' out in a narrow fringe. This fringe is then swept around by
rotating the interferometer, allowing the much fainter planet to be
detected.

An optical telescope produces an image of an unresolved star that
forms the familiar Airy pattern which consists of a bright central peak
surrounded by concentric rings of ever decreasing brightness. These
rings are due to diffraction from the sharp circular aperture of the
telecope. Apodization is a technique that can be used to decrease
brightness of the diffraction rings. It can be accomplished by either
altering the shape or by modifying the transmission of the aperture, or
both. 

In this paper we consider a simpler approach to TPF that achieves the
dynamic range required to separate the light reflected from an
Earth-like planet from that of its star. We suggest that apodizing a
telescope with a filter having an {\it optimized shape and
transmission} would sufficiently reduce the diffraction lobes of the
telescope to allow detection of Earth-like planets orbiting solar-like
stars. Techniques for reducing optical diffraction lobes have been in
use since the 19th century. Analagous algorithms are commonly used for
sidebandreduction in radio astronomy and electronic signal processing
applications (Harris 1978) and an excellent review of optical
techniques is given in (Jacquinot 1964). Oliver (1975) suggested
apodizing the still-to-be-built Large Space Telescope to allow that
telescope to detect Jupiter-sized exo-planets.  In 1980, Black (1980)
carried out laboratory experiments, showing that apodizing could be
used to separate two point sources that had a $10^4$ ratio and a 1
arcsecond separation, comparable to the magnitude ratio and separation
of Sirius and its white dwarf companion. Watson et al (1991) suggest
that apodization could be combined with a coronagraph in a space-based
telescope to directly detect exo-planets.

\section{Approach}

We propose to use an approach that does not require nulling
interferometry to achieve the dynamic range required for planet
detection. This approach provides sufficient dynamic range at
separations of a few times the diffraction limit of a telescope at
visible or near-IR wavelengths, so telescope size (or interferometer
baseline) is greatly reduced.  Our approach is elegantly simple. We
place a square, transmission weighted apodizing aperture in the
telescope pupil or at a relay image of the telescope pupil. The
advantage of a square aperture was suggested by Zanoni and Hill
(1965) who were interested in measuring the gravitational deflection
of starlight by the Sun without having to wait for a solar
eclipse. The square aperture results in most of its extended
diffraction occuring along the x and y axes perpendicular to the edges
of the square and strong suppression around the diagonal directions.
We then apodize the aperture with a transmission function that has
peak transmission in the center of the aperture and drops off toward
the edges of the aperture. This produces extended regions
around the diagonals of the point spread function (PSF) that drop
precipitously in intensity. In those regions, the intensity is less
than $10^{-9}$ of the PSF peak, even for realistic levels of random
wavefront error in the pupil (~1/60 wave of random roughness in the
surface finish) and bandpasses (3000 angstroms). However, it is
obvious that any telescope that can directly image exo-planets will
have to be of extremely high quality. Ripple and imperfections in the
mirror, spiders, or mirror supports will all diffract light at angles
comparable to the expected angular separation of a planet (Angel et
al, 1984, Breckenridge et al, 1984). Trauger et al (2000) suggest
that high order, very small throw adaptive optics may be needed to
correct imperfections in the mirror.

The format of the transmission function is critical to achieving the
full dynamic range. That transmission function is formed by taking the
product of two functions, one aligned with the vertical edges of the
square aperture, and the other aligned with the horizontal
edges. Using these crossed 1-D functions provides 3 to 4 orders of
magnitude better suppression of the diffraction sidelobes than using a
circular symmetric apodization, even when it is combined with a square
aperture.

The intensity of the PSF of a diffraction-limited telescope is just the
square modulus of the Fourier transform of the pupil function of the
telescope. A circular aperture having a constant transmission
inside the pupil and zero outside has a PSF in the familiar Airy pattern
of a bright central peak surrounded by alternate dark and bright rings.

For an unapodized rectangular aperture of width $w$ and height $h$,

\begin {equation}
 I(x,y) = I(0,0) \left( \frac{sin\alpha}{\alpha} \right)^2 \left(
 \frac{sin\beta}{\beta} \right)^2
\end{equation}

where  $\alpha = \frac{\pi x w}{\lambda R}$ , \hspace{0.1in} $\beta = \frac{\pi y h}{\lambda R}$, \hspace{0.1in} and $R =$ distance from pupil to image.

Along the diagonal ($|x| = |y|$) of a square aperture ($h = w$) we get

\begin {equation}
 I(x,x) = I(0,0) \left( \frac{sin\alpha}{\alpha} \right)^4
\end{equation}

The intensity in the diffraction pattern drops rapidly along the
diagonal directions. If we also apodize the square with crossed
transmission functions perpendicular to the edges of the square, the
intensity in the diagonal drops even faster with increasing distance
from the central peak. The transmission of the aperture must have the
general form $T(x,y) = F(x)F(y)$, providing rapid attenuation along
the square aperture diagonal.

We have numerically tested the combination of a square aperture and
several different transmission functions, and the results appear to be
quite insensitive to the functional form of the apodization. The
(slightly) best transmission functions tested so far are the so-called
Sonine functions suggested by Oliver (1975). Crossed Sonine
functions have a transmission function of the form

\[ T(x,y) = \left\{ \begin{array}{ll}
            (1 - x^2)^{\nu-1} (1 - y^2)^{\nu-1}  
         & \mbox{if $-1 \le x \le 1; -1 \le y \le 1$} \\
       0 & \mbox{otherwise}
                \end{array}
                \right. \]
where $\nu$ is a small integer such as 3,4, or 5.

One tradeoff with apodization is that the overall aperture
transmission is reduced below 20\%, so the optimum apodizer is one
that combines the highest transmission with the best reduction of
diffraction sidelobes.  We tested the case of combining a square
aperture and a circularly symmetric transmission Sonine function, but
this results in $10^3$ less dynamic range.  We also tried various values
for $\nu$ with the Sonine function and two other
transmission functions: $cos^2(x) cos^2(y)$ and $cos^4(x) cos^4(y)$
(evaluated from $-\pi/4$ to $\pi/4$). Comparisons of using the different
shaped functions are shown in our simulation results in the next
section.

\section{Simulations}

The dynamic range achieved by using a square aperture combined with
crossed transmission functions is demonstrated by results shown in
Figures 1 - 6.

Figure \ref{fig1} shows the square aperture embedded in a larger field
(upper left): the image of two point sources (100:1 ratio) through the
square aperture with no apodization (upper right); the (Sonine, $\nu$
= 4) apodized aperture (lower left); and the image of the same two
point sources through the apodized aperture (lower right). The images
of the point sources have been expanded around the region where the
sources are located. The point source images all are displayed using
the 4th root of the intensity, scaled up with a gamma of 0.25, using
an IDL color display table. Obviously, the apodization has
dramatically reduced the diffraction sidelobes. For the cases shown
here, the ``planet'' is separated from the star by 6 times the
diffraction-limited element set by the aperture size.

\placefigure{fig1}

Figure \ref{fig2} shows four images through the Sonine apodized
aperture ($\nu$ = 5), again expanding the region around the point
sources. The upper left is for $10^6$ brightness ratio of star to
planet; upper right is $10^9$ ratio of star to planet, no wavefront
aberration; lower left has random error of 1/72 wave rms in the
aperture with $10^9$ ratio; and lower right is 1/50 wave rms random
error, again with a $10^9$ ratio of star to planet.  Very high optical
quality is obviously required to detect Earth-like planets.

\placefigure{fig2}

For comparison, in Figure \ref{fig3}, we show the equivalent imaging
of a pair of point sources through a circular aperture, with and
without Sonine ($\nu = 5$) apodization. The upper two images are for a
100:1 ratio of ``star'' to ``planet''. The lower images are for $10^9$
ratio of ``star'' to planet. Obviously, apodization of a circular
aperture does not come close to providing the dynamic range required
for Earth-like planet detection.

\placefigure{fig3}

Figure \ref{fig4} consists of four contour plots of the region around
the point sources, for different apodizing functions. All plots are
for the case where the point source ratio is $10^9$ and the aberration
is about 1/72 wave.  The contours are on a logarithmic scale with the
lowest contour at $10^{-10}$ and the peak of the "star" at $10^0$. The
two upper plots use crossed Sonine functions for apodization with $\nu
= 4$ (upper left) and $\nu = 5$ (upper right). The lower plots use
crossed $cos^2$ apodization (lower left) and $cos^4$ (lower right).

\placefigure{fig4}

There are small differences in the results. The $\nu = 5$ and $cos^4$
images are slightly cleaner but the transmission of the apertures is
less, as noted on the plots. These results show that the dynamic range
is only a weak function of the specific apodizing shape. They also
show that there is a very large region outside the central diffraction
spikes that has sufficient dynamic range to detect planets, even
terrestrial-sized planets.

Figure \ref{fig5} shows diagonal cuts through the monochromatic PSF
for four cases: 1) circular aperture, no apodization; 2) square
aperture, no apodization; 3) circular aperture with Sonine
apodization; 4) square aperture with crossed Sonine apodization.

\placefigure{fig5}

The results in figures 1-5 were all calculated for monochromatic
light.  In figure \ref{fig6}, we show cuts through the diagonals of the PSF's
for the case where we have integrated over a 3000 $\AA$ bandpass. This
has the effect of reducing the deep nulls in the diffraction patterns.
The vertical dashed line shows the position of a point in the cuts that is 
separated from the central peak by three diffraction-limited elements.
Curve 1 shows a cut through the PSF for an unapodized circular
aperture. Curve 2 shows the effect of Sonine apodization on the PSF
for the circular aperture. Curve 3 shows the apodized square aperture
(Sonine apodization) without the planet (down $10^9$ from the star)
and curve 4 has a planet down $10^9$ from the star. Again, these are
calculated with about 1/72 wave rms aberration in the pupil. Broadening
the bandpass seems to enhance, rather than degrade, the performance
of the apodizers.

\placefigure{fig6}

\section{Discussion}

These results clearly demonstrate the power of the square aperture
when combined with crossed transmission function apodization. One can
apodize in a relay plane of the pupil, allowing use of any shape
telescope aperture initially. The apodizing mask can then be rotated
to sweep out a full field, rather than having to rotate the telescope
or telescope array.  The technique could be combined with the
hypertelescope (Labeyrie 1996, Guyon 1999) approach to building a
large telescope and apodizing the densified pupil.  Because of the
simplicity of the technique, one could easily imagine that it could be
used on a precursor mission to TPF. It might be possible to detect
Earth-like planets around the nearest stars with only a modest
aperture, operating at visible or near infrared wavelengths. At a
minimum, giant planets could be imaged and their spectra measured.

For all of the examples that we have shown, we have located the
``planet'' six diffraction-limited elements from the ``star'', along
the diagonal, where it is well separated. If we move the ``planet'' in
closer to the ``star'', the minimum separation where the ``planet''
signal is just above the wings of the ``star'' diffraction is about
three diffraction limited elements (this may be deduced from the plots
in Figure 6). One can extrapolate this scale to the minimum sizes of
telescopes needed to separate out planets from stars for TPF. For
example, if we wish to image in the wavelength region around $7600
\AA$ where the Oxygen ``A'' line is located, then the diffraction-
limit of a two meter aperture telescope with a square aperture (given
by $\lambda/D$, where D is the telescope diameter) is 0.076
arcsec. Since we can just detect an Earth-like planet spaced three
times the diffraction limit from the star, this would allow detection
of an Earth in a 1 AU orbit at a little over 4 parsecs. Obviously,
planets in larger radius orbits or larger planets could be detected
around more distant stars. Using larger apodized telescopes would
increase the distance and therefore the number of Sun-like stars that
one could survey for planets in 1 AU or larger orbits.

The nearby (3.2 pc) star Epsilon Eridani has recently (Hatzes et al)
been reported to have a Jupiter-like planet orbiting it. An earth-like
planet in a 1 AU orbit around Epsilon Eridani imaged by a 2 meter
apodized telescope would be separated by .3 arcsecond.  2200 photons
would be detected from the planet in one hour for a 3000 $\AA$
bandwidth, 90 \% q.e. detector and 15 \% throughput for the apodized
telescope.  In the same integration time, the Jupiter-like planet,
orbiting at 3.3 AU would give 275000 detected photons (assuming it has
a similar albedo to Jupiter).

There are many additional areas to be explored. Can we combine square
apodization with nulling, or with coronagraphy to allow detection of
planets in closer to the stellar peak, thereby reducing the required
aperture size for detecting planets around more distant stars? What is
the optimum apodization function, making the narrowest diffraction
spikes or maximizing the transmission? Can we combine apodization with
dilute array (interferometric) telescope concepts, in place of a large
monolithic telescope?  What is the best layout for a telescope array?
Can we build a single structure to rigidly support the array?  What
are the detailed specifications on optical quality, at all spatial
scales? Finally, if apodization allows operation at shorter
wavelengths, can spectra at these wavelengths definitively detect the
signatures of life?

We plan to carry out lab tests to demonstrate the effectiveness of
apodization techniques. Relatively straightforward experiments using
superpolished spherical mirrors and pairs of point sources with a
controlled flux ratio should allow feasibility testing of this
approach.

\section{Acknowledgements}
The authors would like to acknowledge very helpful suggestions and
discussions with Steven Ridgway, Robert Stachnik, Daniel Gezari and
Rick Lyon.  We also wish to acknowledge support for the work from an
SVS subcontract (SVS Inc SUB-00-095) for TPF studies to the
Smithsonian Astrophysical Observatory.

\clearpage

\figcaption[fig1.eps]{
Image of two point sources through apodized and unapodized square 
apertures.  
\label{fig1}}

\figcaption[fig2.eps]{
 Image of two point sources through an apodized
and unapodized square aperture. Upper left: $10^6$ point source ratio;
Upper right: $10^9$ ratio, 1/100 wave rms wavefront error; Lower Left:
$10^9$ ratio and 1/72 wave rms wavefront error; Lower right: $10^9$
ratio and 1/50 wave rms wavefront error.  
\label{fig2}}

\figcaption[fig3.eps]{
Image of two point sources through an apodized and unapodized
circular aperture. Upper left: 100:1 point source ratio, no
apodization; Upper right: 100:1 ratio with Sonine apodization ($\nu =
5$); Lower Left: $10^9$ ratio, no apodization; Lower right: $10^9$
ratio with Sonine apodization.
\label{fig3}}

\figcaption[fig4.eps]{
Contour plots of two point sources with $10^9$ ratio and 1/72
wave rms wavefront error imaged with a square aperture and different
apodization. Upper left: Sonine apodization, $\nu=4$; Upper right:
Sonine apodization, $\nu=5$ ; Lower Left: $Cos^2$ apodization; Lower right:
$Cos^4$ apodization
\label{fig4}}

\figcaption[fig5.eps]{
Diagonal cuts through monochromatic PSF's. 1) Unapodized circular
aperture; 2) Square aperture, no apodization; 3) Circular aperture, Sonine
apodization ($\nu$ = 4); 4) Square aperture, Sonine apodization.  
\label{fig5}}

\figcaption[fig6.eps]{ Diagonal cuts through PSF images for
polychromatic light. 1) Circular aperture, no apodization; 2) Circular
aperture, Sonine apodization; 3) Square aperture, Sonine apodization;
4) Square aperture, Sonine apodization, plus planet down $10^9$ from
the star. The vertical dashed line shows the point in the cut that is
separated by three diffraction-limited elements from the central peak.
\label{fig6}}

\end{document}